\documentclass[aps,prd,amsmath,amssymb,showpacs]{revtex4}
\usepackage{graphicx,color,dcolumn}
\usepackage{amsmath}
\usepackage{graphicx}
\usepackage{subfigure}

\begin{document}
\title{Production of $Y(4260)$ as a hadronic molecule state of $\bar{D}D_1 +c.c.$ in $e^+e^-$ annihilations }

\author{Wen Qin$^1$\footnote{{\it Email address:} qinwen@ihep.ac.cn}, Si-Run Xue$^1$\footnote{{\it Email address:} xuesr@ihep.ac.cn}, Qiang Zhao$^{1,2}$\footnote{{\it Email address:} zhaoq@ihep.ac.cn}}

\affiliation{ 1) Institute of High Energy Physics and Theoretical Physics Center for Science Facilities,
        Chinese Academy of Sciences, Beijing 100049, China}

\affiliation{ 2) Synergetic Innovation Center for Quantum Effects and Applications (SICQEA), Hunan Normal University,Changsha 410081, China}

\date{\today}

\begin{abstract}

We study the $Y(4260)$ production mechanism in $e^+e^-$ annihilations in the framework of hadronic molecules and investigate the consequence of such a picture in different decay channels. In the hadronic molecule picture the $Y(4260)$ is described as a mixture state composed of a long-ranged $\bar{D}D_1(2420)+c.c.$ molecule state and a compact $c\bar{c}$ component. We show that the compositeness relation can still provide a reasonable constraint on the wavefunction renormalization parameter due to the dominance of the molecular component. Such a mechanism can be regarded as a natural consequence of the heavy quark spin symmetry (HQSS) breaking. This study elaborates the molecular picture for the $Y(4260)$ in the $e^+e^-$ annihilations and  affirms that the cross section lineshape of $e^+e^-\to \bar{D}D^*\pi+c.c.$ in the vicinity of the $Y(4260)$ should have a nontrivial behavior. In this framework we predict that the upper limit of the $Y(4260)$ leptonic decay width is about 500 eV. We also investigate the coupling for $D_1(2420)\to D^*\pi$ in the $^3P_0$ quark model and examine the possible HQSS breaking effects due to the deviation from the $|^1P_1\rangle$ and $|^3P_1\rangle$ ideal mixing. This will in turn provide a constraint on the HQSS breaking coupling for the $Y(4260)$ to $\bar{D}D_1(2420)+c.c.$ via its $c\bar{c}$ component.

\end{abstract}


%12.39.Hg Heavy quark effective theory
%14.40.Lb Charmed mesons (|C|>0, B=0)
%14.40.Rt Exotic mesons
%13.66.Bc Hadron production in e?e+ interactions

\maketitle

\section{Introduction}

The study of hadron spectroscopy has made significant progress during the past decade benefiting from the experimental observations of a lot of new states in various processes. In particular, in the heavy flavor sector, a large number of the so-called $X$, $Y$, $Z$ states have been candidates for exotic hadrons which may contain more complicated quark-gluon structures than the conventional quark model picture. Among all those exotic candidates, the $Y(4260)$ is undoubtedly one of the most mysterious states and has initiated a lot of experimental and theoretical studies since its observation by BaBar Collaboration in the $J/\psi\pi\pi$ channel in 2005~\cite{Aubert:2005rm}. It was later confirmed by Belle~\cite{Yuan:2007sj} and CLEO-c Collaboration~\cite{Coan:2006rv} in the same decay channel, but ``evaded" from observation in its decays into open charm channels~\cite{Pakhlova:2009jv} which is unexpected for conventional charmonium states above the $D\bar{D}$ threshold. Nevertheless, in the $R$ value measurement the inclusive cross sections for $e^+e^-$ annihilations appear to have a dip instead of a peak in the vicinity of the $Y(4260)$ mass region. Another important observation which makes $Y(4260)$ peculiar is that the successful potential quark model calculations did not predict a vector state in such an energy region and the conventional $c\bar{c}$ states have been assigned to other structures~\cite{Eichten:1978tg,Eichten:1979ms,Godfrey:1985xj}.

Such surprising properties have initiated a lot of theoretical interpretations of its nature in the literature, such as vector hybrid candidate Refs.~\cite{Zhu:2005hp,Kou:2005gt,Close:2005iz}, hadro-quarkonium~\cite{Voloshin:2007dx,Dubynskiy:2008mq}, tetraquark state~\cite{Maiani:2005pe}, and hadronic molecules of $\bar{D}D_1(2420)+c.c.$~\cite{Ding:2008gr}, $\omega\chi_{c0}$~\cite{Dai:2012pb}, or $J/\psi K\bar{K}$~\cite{MartinezTorres:2009xb}, or non-resonance explanation due to state interferences~\cite{Chen:2010nv}. In Ref.~\cite{LlanesEstrada:2005hz} $Y(4260)$ was proposed to be a conventional charmonium $\psi(4S)$ state. Interestingly, the Lattice QCD (LQCD) calculation indicates that the hybrid vector charmonium is located in the mass region near 4.3 GeV which makes the $Y(4260)$ a possible hybrid state corresponding to the LQCD spectrum~\cite{Liu:2012ze}. In the recent LQCD study of Ref.~\cite{Chen:2016ejo}  a spatially extended hybrid-like operator is applied and yields a rather small leptonic decay width of less than 40 eV. It makes the measurement of the leptonic decay width an interesting observable for probing its internal structures.

The recent observation of the charged charmonium states $Z_c(3900)$ at the mass of $Y(4260)$ in $e^+e^-\to J/\psi\pi\pi$ with high statistics at BESIII~\cite{Ablikim:2013mio} once again attracted attention from the community and provoked further studies of the $Y(4260)$. The mass of the $Z_c(3900)$ is close to the $\bar{D}D^*+c.c.$ threshold. With $J^P=1^+$ it decays via an $S$ wave into $\bar{D}D^*+c.c.$ and $J/\psi\pi$ with large coupling to $\bar{D}D^*+c.c.$ which makes it a reasonable candidate for hadronic molecule of $\bar{D}D^*+c.c.$ Nevertheless, its production from the $Y(4260)$ decays provides important information for the structure of $Y(4260)$ and a detailed studies of the $Y(4260)$ should be crucial for a self-consistent picture for those puzzling phenomena arising in this energy region.

In Ref.~\cite{Wang:2013cya}, it was proposed that the production of $Z_c(3900)$ in $Y(4260)\to J/\psi\pi\pi$ should be a strong signal for $Y(4260)$ being a $\bar{D}D_1(2420)+c.c.$ hadronic molecule. One notices that the $\bar{D}D_1(2420)+c.c.$ is the first two-body open charm $S$-wave threshold in the vector charmonium channel with narrow final states and the $Y(4260)$ is less than 30 MeV below the $\bar{D}D_1(2420)+c.c.$ threshold. It should be pointed out that the mass and width of $Y(4260)$ were extracted by a Breit-Wigner fit to the $e^+e^-\to J/\psi\pi\pi$ cross sections~\cite{Agashe:2014kda}. Since the cross section lineshape is not symmetric at the two sides of the peak this treatment may cause significant uncertainties with the resonance parameters. In fact, the asymmetric lineshape can be regarded as an indication of the nontrivial nature of the $Y(4260)$ of which the propagator cannot be parameterized by a simple Breit-Wigner. This forms one of the motivations in this study that a proper treatment of the renormalization effects due to the meson loops will be essential for extracting the pole information.

Recall that we have been emphasizing that a strong $S$-wave interactions between $\bar{D}D_1(2420)+c.c.$ can lead to a dynamically generated pole structure near the $\bar{D}D_1(2420)+c.c.$ threshold and the $Y(4260)$ should be the natural candidate~\cite{Ding:2008gr,Wang:2013cya}. To firm up such a scenario, various properties and observables were investigated in the molecular picture. In Ref.~\cite{Guo:2013nza}, it was studied that the $Y(4260)$ as a $\bar{D}D_1(2420)+c.c.$ molecule will have sizeable decay rate into $\gamma X(3872)$ given that $X(3872)$ is also a hadronic molecule of $\bar{D^0}D^{*0}+c.c.$ One interesting question arising from the $\bar{D}D_1(2420)+c.c.$ molecule picture is that the $D_1(2420)$ should be the narrow state in the $^3P_1$ and $^1P_1$ mixing for the $D_1$ states. The data show that the narrow $D_1(2420)$ will be dominated by the configuration with the light quark angular momentum $(1^- + 1/2^-=3/2^+)$ coupled by the orbital angular momentum and antiquark spin and parity. This will be a pure configuration either by taking the heavy quark spin symmetry (HQSS) limit or if $D_1(2420)$ is one of the eigenstate from the  $^3P_1$ and $^1P_1$ ideal mixing. It was then pointed out in Ref.~\cite{Li:2013yka}, in the HQSS limit the coupling of the heavy $c\bar{c}$ in an $S$ wave of $^3S_1$ while the light quark degrees of freedom with the spin and orbital angular momentum coupling to $J^P=0^+$ is forbidden in the spin decomposition of $\bar{D}D_1(2420)+c.c.$ This observation brings concerns on the production of $Y(4260)$ in $e^+e^-$ annihilations if it is indeed a hadronic molecule of $\bar{D}D_1(2420)+c.c.$ However, as shown by Refs.~\cite{Cleven:2013mka,Wang:2013kra}, the HQSS breaking in the charmonium mass region should be expected and it is sufficient to account for the cross sections observed for $Y(4260)$. Our studies in this work will further clarify the HQSS breaking effects and make the molecular picture dynamically self-contained.

In brief, we will follow the molecular picture of Refs.~\cite{Wang:2013cya,Guo:2013nza,Cleven:2013mka,Wang:2013kra} and explore the $Y(4260)$ production mechanism in $e^+e^-$ annihilations. We will try to accommodate the short-distance production mechanism coherently in the molecular picture using Weinberg's compositeness theorem~\cite{Weinberg:1965zz,Baru:2003qq}. This will quantify the HQSS breaking effects in explicit calculation and this treatment should be able to tell us more about the structure of the $Y(4260)$.

As follows, we first present the full propagator of the $Y(4260)$ in the framework of the non-relativistic effective field theory (NREFT) in Sec.~\ref{nreft-propagator}. In Sec.~\ref{lineshape-study} we present the formalism for the study of the cross section lineshape of $e^+e^-\to \bar{D}D^*\pi+c.c.$ and the width of $Y(4260)\to e^+e^-$. We then investigate the HQSS breaking effects in the $^3P_0$ model as an independent check of the HQSS breaking in Sec.~\ref{3P0-model}. A brief summary is given in Sec.~\ref{summary}.

\section{The molecular picture for $Y(4260)$}\label{nreft-propagator}

\begin{figure}[htb]
\centering
\scalebox{0.6}{\includegraphics{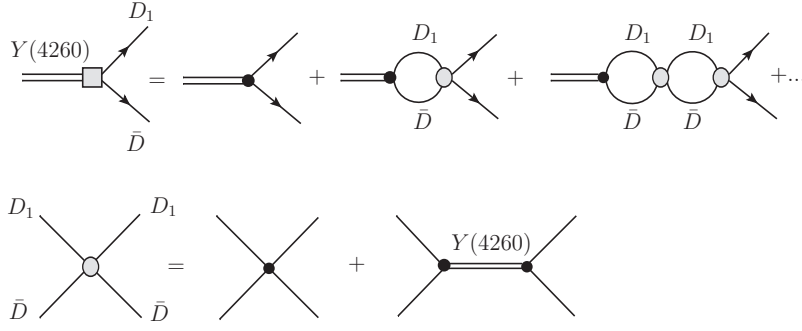}}
\caption{The bubble diagrams for the coupling of $Y(4260)$ to $\bar{D}D_1+c.c.$ They rescattering vertex (shadowed circle) contains the contact interaction and bare propagator of the $Y(4260)$. } \label{loop}
\end{figure}

In our scenario $Y(4260)$ is treated as an $S-$wave molecule of $\bar{D}D_1+c.c.$ containing a small charmonium component $|c\bar{c}\rangle$, i.e.
\begin{equation}\label{decomp}
|Y(4260)\rangle=\alpha|c\bar{c}\rangle+\beta|\bar{D}D_1+c.c.\rangle,
\end{equation}
where the component strength satisfies the normalization relation $\alpha^2+\beta^2=1$. For the $\bar{D}D_1$ scattering (the conjugate part is implicated), the contact interaction is also introduced in addition to the bare pole structure. The process for  the decay of $Y(4260)\to \bar{D}D_1+c.c.$ and scattering of $\bar{D}D_1\to \bar{D}D_1$  are illustrated in Fig.~\ref{loop}. The corresponding Lagrangian can be written as
\begin{eqnarray}\label{lagrangian-YD1D}
\mathcal{L}_Y&=&\frac{y^{\text{bare}}}{\sqrt{2}}(\bar{D}_a^\dagger Y^iD^{i\dagger}_{1a}-\bar D^{i\dagger}_{1a} Y^iD^\dagger_a)+g_1\{(D^i_{1a}\bar{D}_a)^\dagger(D^i_{1a}\bar{D}_a)+(D_{a}\bar{D}^i_{1a})^\dagger(D_{a}\bar{D}^i_{1a})\}+H.c. ,
\end{eqnarray}
where the real coefficient $g_1$ is the coupling constant of the contact interaction.

In the molecular scenario the non-relativistic propagator of the $Y(4260)$ can be expressed as
\begin{eqnarray}\label{propagator}
\mathcal{G}_Y(E)&=&\frac{1}{2}\frac{i}{E-m_0+\Sigma_{\bar{D}D_1}(E)\times[i(y^\text{bare})^2-4i(E-m_0)g_1]}\nonumber\\
&\equiv & \frac{1}{2}\frac{i}{E-m_0-\Sigma_1(E)} ,
\end{eqnarray}
where $m_0$ is the bare mass of $Y(4260)$ and $\Sigma_1(E)\equiv\Sigma_{\bar{D}D_1}(E)(-i(y^\text{bare})^2+4i(E-m_0)g_1)$ comes from the sum of the infinite bubble loops in the $\bar{D}D_1$ rescattering. The one loop function is defined as the following:
\begin{eqnarray}
\Sigma_{\bar{D}D_1}(E)&=&\frac{-1}{4}\int\frac{d^Dl}{(2\pi)^D}\frac{1}{(l^0-m_{D}-\overrightarrow{l}^2/(2m_{D})+i\epsilon)(E-l^0-m_{D_1}-\overrightarrow{l}^2/(2m_{D_1})+i\Gamma_{D_1}/2)} .
\end{eqnarray}
In dimensional regularization with the $\overline{\text{MS}}$ subtraction scheme, the loop integral $\Sigma_{\bar{D}D_1}(E)$ is finite for $D=4$ and one obtains
\begin{eqnarray}
\Sigma^{\overline{\text{MS}}}_{\bar{D}D_1}(E)&=&\frac{\mu}{8\pi}\sqrt{2\mu(E-m_D-m_{D_1})+i\mu\Gamma_{D_1}},
\end{eqnarray}
where $\mu=m_D m_{D_1}/(m_D+m_{D_1})$ is the reduced mass. The bare mass $m_0$ should be renormalized to the physical mass $m_Y$, which gives the physical pole position in the real axis of the energy plane via $m_Y=m_0 + \text{Re}\Sigma_1(m_Y)$.

By expanding the denominator of the propagator near the physical pole position of $m_Y$, the propagator can be expressed as
\begin{eqnarray}
\mathcal{G}_Y(E)&=&\frac{1}{2}\frac{iZ}{E-m_Y-Z\widetilde{\Sigma}_1(E)} ,
\end{eqnarray}
where $\widetilde{\Sigma}_1(E)\equiv {\Sigma}_1(E)-\text{Re}(\Sigma_1(m_Y))-(E-m_Y)\text{Re}(\partial_E\Sigma_1(m_Y))$ and $Z\equiv 1/[1-\text{Re}(\partial_E\Sigma_1(m_Y))]$ is the wavefunction renormalization constant. Note that $(1-Z)$ is the probability of finding $Y(4260)$ in a $\bar{D}D_1+c.c.$ molecular state. It can then be related to the molecular component of the $Y(4260)$ wavefunction via  $|\beta|=\sqrt{1-Z}$. Alternatively, the renormalization constant $Z$ defines the probability of finding the $Y(4260)$ in a non-molecular state. A natural assumption is that $Z$ can be related to the compact $c\bar{c}$ component in the $Y(4260)$ wavefunction and at leading order one expects $|\alpha|\simeq \sqrt{Z}$. It should be emphasized that this renormalization will allow  constraints on the production of $Y(4260)$ via the short-distance component, e.g. in the $e^+e^-$ annihilations. This also implies that $\bar{D}D^*\pi+c.c.$ should be one of the important decay channels of $Y(4260)$. However, notice that the $D$-wave decay of $D_1(2420)\to D^*\pi$ is suppressed. The partial width of  $Y(4260)\to \bar{D}D^*\pi+c.c.$ is not expected to be sizeable. A detailed study is necessary for a coherent picture for the nature of $Y(4260)$.

We also introduce a constant width $\Gamma^{\text{non}-\bar{D}D_1}$ in the propagator to account for contributions from decay channels other than the $\bar{D}D_1+c.c.$
Then the complete propagator of $Y(4260)$ reads
\begin{eqnarray}\label{propagator-full}
\mathcal{G}_Y(E)&=&\frac{1}{2}\frac{iZ}{E-m_Y-Z\widetilde{\Sigma}_1(E)+i\Gamma^{\text{non}-\bar{D}D_1}/2},
\end{eqnarray}
where $m_Y$ and $\Gamma^{\text{non}-\bar{D}D_1}$ are parameters to be fitted by experimental data for the cross section lineshapes of $e^+e^-\to J/\psi\pi\pi$ and $h_c\pi\pi$. As studied in Ref.~\cite{Cleven:2013mka}, the asymmetric cross section lineshape of $e^+e^-\to J/\psi\pi\pi$ turns out to be selective for the determination of the pole position $m_Y$ and $\Gamma^{\text{non}-\bar{D}D_1}$. Thus, by fitting the cross section lineshapes of $e^+e^-\to J/\psi\pi\pi$ and $h_c\pi\pi$, but allowing an overall scale parameter, we can determine $m_Y=4.217\pm0.002$ GeV and $\Gamma^{\text{non}-\bar{D}D_1}=0.056\pm0.003$ GeV~\cite{Cleven:2013mka}. We note that the pole corresponding to $Y(4260)$ is close to the threshold of $\bar{D}D_1+c.c.$ when considering the pole value of $D_1(2420)$.  
This follows the approach of Ref.~\cite{Cleven:2013mka}, namely, the renormalization constant $Z$ in the numerator of Eq.~(\ref{propagator-full}) was absorbed into the vertex coupling constant as an overall scale parameter. This is because that the pole is fitted at $E-m_Y-Z\text{Re}\widetilde{\Sigma}_1(E)=0$ with $E=m_Y$, i.e. $\text{Re}\widetilde{\Sigma}_1(m_Y)=0$. It means that the pole position does not explicitly depend on the renormalization constant.

One notices that $\Gamma^{\text{non}-\bar{D}D_1}=0.056\pm0.003$ GeV accounts for the sum of contributions from non-$\bar{D}D_1+c.c.$ channels. They include the hidden charm decays, i.e. $J/\psi\pi\pi$, $h_c\pi\pi$, and $\omega\chi_{c0}$, and also open charm decays that do not go through the $\bar{D}D_1+c.c.$ This part of open charm decays reflect the HQSS breaking effects for which we will try to quantify in this work later.

It should be noted that the recent experimental measurements of BESIII~\cite{Ablikim:2014qwy,Ablikim:2015uix} suggest that the cross sections for $e^+e^-\to J/\psi\pi\pi$, $h_c\pi\pi$, and $\omega\chi_{c0}$ are compatible. If these channels saturate the total width of $Y(4260)$, it will lead to relatively small leptonic decay width for $Y(4260)\to e^+e^-$~\cite{Dai:2012pb}. This is also a challenge in understanding the nature of $Y(4260)$.

As follows, we will combine the production mechanism to investigate the relation between the renormalization constant $Z$ and its implication of relative decay strength for $Y(4260)$ to different decay channels.  The full propagator of Eq.~(\ref{propagator-full}) and decomposition of the $Y(4260)$ wavefunction of Eq.~(\ref{decomp}) will help us to separate out the short-distance compact $c\bar{c}$ component and long-distance molecular component, and also provide constraint on its leptonic decay width.

\section{The line shapes of $e^+e^- \to Y(4260) \to \bar{D}D^*\pi+c.c.$ and the width of $Y(4260)\to e^+e^-$}\label{lineshape-study}

\begin{figure}[htb]
\centering
\scalebox{0.6}{\includegraphics{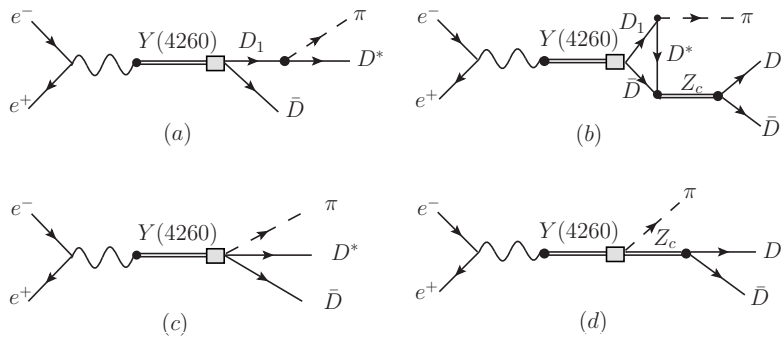}}
\caption{Feynman diagrams for $e^+e^- \to Y(4260) \to \bar{D}D^*\pi+c.c.$ via the molecular component (a) and (b) and compact $c\bar{c}$ component (c) and (d). The production of $Z_c(3900)$ is included. } \label{Y-decay}
\end{figure}

The production and decay of $Y(4260)$ in $e^+e^-$ annihilations can go through  two steps, i.e. the compact $c\bar{c}$ component is first created by the virtual photon and then the physical $Y(4260)$ can decay either via the molecular $\bar{D}D_1+c.c.$ component or the compact $c\bar{c}$ into final $\bar{D}D^*\pi+c.c.$ It can then be illustrated by Fig.~\ref{Y-decay} where (a) and (b) are decay processes through the molecular component while (c) and (d) are through the compact $c\bar{c}$ which are originated from the HQSS breaking.

Although there are various interpretations for the $Z_c(3900)$, we treat it as
an $S-$wave molecule of $\bar{D}D^*$ in our framework. Similar to $Y(4260)$ the propagator of the $Z_c(3900)$ can be written as
\begin{eqnarray}\label{Zcprop}
\mathcal{G}_{Z_c}(E)&=&\frac{1}{2}\frac{i}{E-m_Z+\Sigma_{\bar{D}D^*}(E)\times[iz^2-2i(E-m_Z)g_2]+i\Gamma^{\text{non}-\bar{D}D^*}/2} \ ,
\end{eqnarray}
with
\begin{eqnarray}
\Sigma_{\bar{D}D^*}(E)&\equiv &\frac{\mu^\prime}{8\pi}(\sqrt{2\mu^\prime\epsilon}\theta(\epsilon)+i\sqrt{-2\mu^\prime\epsilon}\theta(-\epsilon)),
\end{eqnarray}
where $\mu^\prime=m_D m_{D^*}/(m_D+m_{D^*})$, $\epsilon=E-m_D-m_{D^*}$ and $|z|\approx0.77$ GeV$^{-1/2}$ is the coupling constant of $Z_c(3900)$ to $\bar{D}D^*$~\cite{Cleven:2013mka}. The contact interaction of $\bar{D}D^*\to \bar{D}D^*$ is introduced by the term with coupling $g_2$ in Eq.~(\ref{Zcprop}). However, by fitting the $\bar{D}D^*+c.c.$ invariant mass spectrum it shows that the value of $g_2$ is negligibly small and so we take $g_2=0$ in the following analysis.

Given that a compact $c\bar{c}$ component is present in the wavefunction of $Y(4260)$, it would be a natural mechanism to produce the $Y(4260)$ in the $e^+e^-$ annihilations. The coupling of $Y(4260)$ to the virtual photon can be approximated by the vector meson dominance (VMD) model, which is described by the following Lagrangian for the compact component coupling to the virtual photon:
\begin{eqnarray}
\mathcal{L}_{Y\gamma}&=&\frac{em^2_Y}{f_Y}Y_\mu A^\mu \ ,
\end{eqnarray}
where $f_Y$ is the bare decay constant defined for the compact wavefunction of $Y(4260)$.

It should be addressed that the compact $c\bar{c}$ in an $S$ wave should decouple to the $\bar{D}D_1+c.c.$ molecular component in the HQSS limit~\cite{Wang:2013cya,Guo:2013nza,Cleven:2013mka,Wang:2013kra}. However, since the charm quark is not heavy enough the HQSS breaking is anticipated and there will be several possible mechanisms leading to mixing between the compact $c\bar{c}$ and molecular $\bar{D}D_1+c.c.$:

\begin{itemize}

\item The presence of the compact $c\bar{c}$ component may be produced by the coupled channel between $\bar{D}D_1+c.c.$ and $\bar{D}^*D_0+c.c.$ which allows a small $S$-wave coupling for $Y(4260)\to \bar{D}^*D_0+c.c.\to \bar{D} D^*\pi +c.c.$ Since the $D_0$ state is very broad such a contribution including the $Z_c(3900)$ pole, i.e. contributions from Fig.~\ref{Y-decay} (c) and (d), can be parameterized by an $S$-wave coupling amplitude for $Y(4260)\to \bar{D} D^*\pi +c.c.$ as demonstrated in Ref.~\cite{Cleven:2013mka}:
\begin{eqnarray}\label{s-wave-para}
A_S&=&a(M^2_{\bar{D}D^*}+b)\mathcal{G}_{Z_c}(E) \ ,
\end{eqnarray}
where $M^2_{\bar{D}D^*}$ is the invariant mass square of $\bar{D}$ and $D^*$ in the final state and $a$ and $b$ are real numbers to be fitted by the experimental data. This parametrization~\cite{Hanhart:2012wi} respects Watson theorem and is equivalent to introducing a contact interaction and a pole term to the transition amplitude as illustrated by Fig.~\ref{Y-decay} (c) and (d), respectively.

\item The coupled channel between $\bar{D}D_1+c.c.$ and $\bar{D}^*D_0+c.c.$ will lead to the HQSS breaking and result in the deviation of the $^3P_1$ and $^1P_1$ mixing from the ideal mixing. Then, a small $S$-wave coupling for $Y(4260)$ to the physical $\bar{D}D_1+c.c.$ will be allowed. This mechanism will also affect the decay of $D_1(2420)\to D^*\pi$ where a small $S$-wave coupling will contribute to the amplitude of Eq.~(\ref{s-wave-para}). This scenario can be examined by the quark model calculation and we will come back to this point later in this work.

\end{itemize}

The above considerations can be recognized by the full amplitude of $e^+e^- \to Y(4260) \to \bar{D}D^*\pi+c.c.$ which is expressed as
\begin{eqnarray}\label{trans-amp}
\mathcal{M}_{\bar{D}D^*\pi}&=&\bar{v}(p_1)(-ie\gamma^i)u(p_2)\frac{i}{E^2}\frac{iem^2_Y}{f_Y}\frac{1}{2}\frac{i Z}{E-m_Y-Z\widetilde{\Sigma}_1(E)+i\Gamma^{\text{non}-\bar{D}D_1}/2}\nonumber \\&&\times
\Big{\{}A_S\delta^{ij}+(\frac{iy^{\text{bare}}}{\sqrt{2}})[\frac{i}{s_1-m^2_{D_1}+im_{D_1}\Gamma_{D_1}}+(-z^2)\mathcal{I}(E,\overrightarrow{p}_\pi^2)\mathcal{G}_{Z_c}(s_2)]
\times\frac{h^\prime}{f_\pi}(3p_\pi^ip_\pi^j-\overrightarrow{p}_\pi^2\delta^{ij})\Big{\}}\epsilon_{D^*}^j\nonumber \\
&\equiv&\bar{v}(p_1)(\gamma^i)u(p_2)\frac{ie^2m^2_Y}{f_YE^2}\frac{1}{2}\frac{iZ}{E-m_Y-Z\widetilde{\Sigma}_1(E)+i\Gamma^{\text{non}-\bar{D}D_1}/2}\nonumber \\&&\times
\Big{\{}A_S\delta^{ij}+(\frac{iy^{\text{bare}}}{\sqrt{2}})A_D(E,M_{\bar{D}D^*},M_{D^*\pi})
\times\frac{h^\prime}{f_\pi}(3p_\pi^ip_\pi^j-\overrightarrow{p}_\pi^2\delta^{ij})\Big{\}}\epsilon_{D^*}^j,
\end{eqnarray}
where $s_1=(p_{\pi}+p_{D^*})^2$, $s_2=(p_{D}+p_{D^*})^2$ and $\mathcal{I}(E,\overrightarrow{p}_\pi^2)$ is the triangle loop integral,
\begin{eqnarray}
\mathcal{I}(E,\overrightarrow{p}_\pi^2)&=&\frac{1}{8}\int\frac{d^4l}{(2\pi)^4}\frac{i}{(l^0-m_{D_1}-\overrightarrow{l}^2/(2m_{D_1})+i\Gamma_{D_1}/2)}\frac{i}{(E-l^0-m_D-\overrightarrow{l}^2/(2m_{D})+i\epsilon)}\nonumber\\&&
  \times\frac{i}{(l^0-p^0_{\pi}-m_{D^*}-(\overrightarrow{l}-\overrightarrow{p}_{\pi})^2/(2m_{D})+i\epsilon)} \ .
\end{eqnarray}
In the above equation, $A_S$ is the $S$-wave strength and $A_D(E,M_{\bar{D}D^*},M_{D^*\pi})$ is the $D$-wave strength via the intermediate $D_1$ while $p_2$ and $p_1$ are the incoming momenta of the electron and positron, respectively. This expression is different from the treatment of Ref.~\cite{Cleven:2013mka} where the wavefunction renormalization constant $Z$ in the numerator has been absorbed into an overall scale parameter. In Eq.~(\ref{trans-amp}) the renormalization constant is explicitly included based on the compositeness theorem. One notices that the effective coupling for $Y(4260)$ to $\bar{D}D_1+c.c.$ and a virtual photon can be renormalized as $y^\text{eff}=\sqrt{Z}y^\text{bare}$ and $\frac{1}{f^\text{eff}_Y}=\sqrt{Z}\frac{1}{f_Y}$, respectively.

The parameters in Eq.~(\ref{trans-amp}) includes the bare coupling $y^\text{bare}$ for $Y(4260)$ to $\bar{D}D_1+c.c.$, bare decay constant $1/f_Y$ for $Y(4260)$ to virtual photon, and  parameters $a$ and $b$ for the small $S$-wave contributions. Note that the renormalization constant $Z$ is a function of $y^\text{bare}$.  For the $Y(4260)$ being a $\bar{D}D_1 +c.c.$ molecule, the value of $Z$ which representing the non-$\bar{D}D_1+c.c.$ component will take the zero limit. It means that the production of $Y(4260)$ in $e^+e^-$ annihilations will be suppressed. Alternatively, if there is a sizeable non-$\bar{D}D_1+c.c.$ component inside $Y(4260)$, its production in $e^+e^-$ annihilations will be enhanced. However, its coupling to $\bar{D}D_1+c.c.$ will be suppressed by the compositeness theorem. Such a constraint can provide a self-consistent check of the molecular scenario.

The $D_1(2420) D^*\pi$ coupling $h'$ can be determined by $D_1(2420)\to D^*\pi$ via the effective coupling~\cite{Cleven:2013mka}:
\begin{eqnarray}\label{lagrangian-D1}
\mathcal{L}_{D_1}&=&i\frac{h^\prime}{f_\pi}\Big[3D^i_{1a}(\partial^i\partial^j\phi_{ab})D^{*\dagger j}_b-D^i_{1a}(\partial^j\partial^j\phi_{ab})D^{*\dagger i}_b-3\bar D^{*\dagger i}_{a}(\partial^i\partial^j\phi_{ab})\bar D^j_{1b}+\bar D^{*\dagger i}_{a}(\partial^j\partial^j\phi_{ab})\bar D^i_{1b}\Big]+H.c.
\end{eqnarray}
Note that the neutral and charged $D_1(2420)$ have similar total widths, i.e. $\Gamma_{D_1^0(2420)}=27.4\pm 2.5$ MeV and $\Gamma_{D_1^\pm(2420)}=25\pm 5$ MeV~\cite{Agashe:2014kda}. Meanwhile, the partial width for $D_1(2420)\to D^*\pi$ has not been precisely measured. Therefore, the total width will give the upper limit of the coupling constant $h'$ if one assumes that the total width of $D_1(2420)$ is saturated by the $D^*\pi$ decay.

\begin{figure}[htb]
\centering
\scalebox{1}{\includegraphics[width=8cm, height=5.5cm]{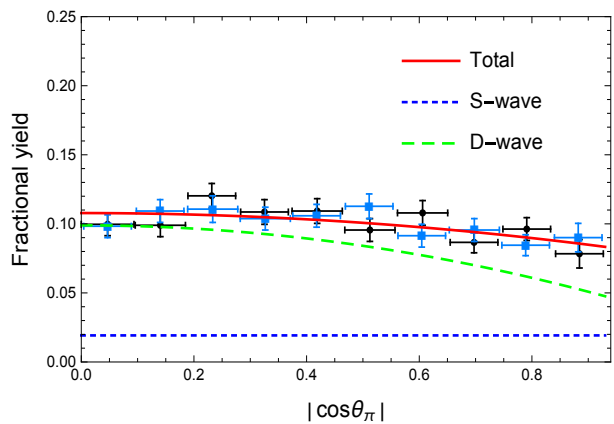}}
\caption{Angular distribution of the pion in the $e^+e^-$ c.m. frame with respect to the beam axis. The recoiled part is the threshold enhancement of $\bar{D}D^*$. The data are from Ref.~\cite{Ablikim:2015swa} with the round dots measured in $\bar{D}D^*+c.c.$ decays into $\pi^+D^0\bar{D}^0$ channel and squared dots measured in the $\pi^+D^0D^-$ channel.  }\label{Fractional}
\end{figure}

\begin{figure}[htb]
\centering
\scalebox{1}{\includegraphics{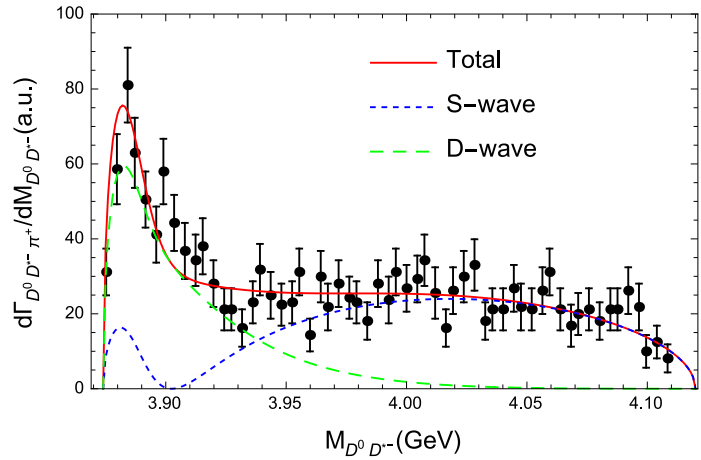}}
\caption{The $D^0D^{*-}$ invariant mass distribution for $Y(4260) \to D^0D^{*-}\pi^+$. The dotted, dashed and solid lines are the contributions from the $S-$wave, $D-$wave and the sum of them, respectively. The data are from Ref.~\cite{Ablikim:2015swa}. }\label{dsigma}
\end{figure}

\begin{figure}[htb]
\centering
\scalebox{1}{\includegraphics{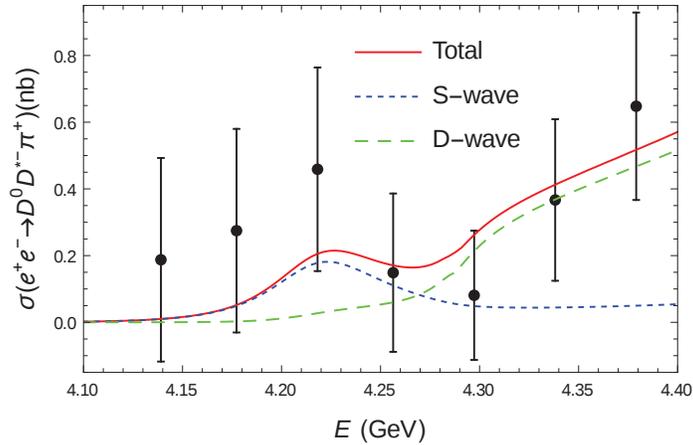}}
\caption{The cross section of $e^+e^- \to Y(4260) \to D^0D^{*-}\pi^+$ with the dotted, dashed and solid lines denoting the contributions from the $S-$wave, $D-$wave and the sum of them, respectively. The data are from Ref.~\cite{Pakhlova:2009jv}.} \label{sigma}
\end{figure}

By applying the pole mass and non-$\bar{D}D_1$ width from Ref.~\cite{Cleven:2013mka}, we fit the BESIII data~\cite{Ablikim:2015swa} for the angular distribution of pion which recoils the $\bar{D} D^*$ threshold peak in the $e^+e^-$ c.m. frame, the invariant mass spectrum of the $D^0D^{*-}$, and the Belle data~\cite{Pakhlova:2009jv} for the cross section for $e^+e^-\to Y(4260)\to D^0D^{*-}\pi^+$ to extract the couplings. The fitting results are illustrated in Figs.~\ref{Fractional}-\ref{sigma}, respectively.
Similar to what found in Ref.~\cite{Cleven:2013mka}, as shown by Fig.~\ref{Fractional} the angular distribution of the recoiled pion provides a strong constraint on the $S$-wave component. A flat angular distribution of the pion recoiling the threshold enhancement in the $\bar{D}D^*$ spectrum can be produced with a dominant $D$-wave and relatively small $S$-wave amplitudes. Actually, the improved data from Ref.~\cite{Ablikim:2015swa} appear to have tilted behavior deviating from a pure $S$-wave angular distribution. This could be an indication of presence of non-$S$ component.

One can see in Fig.~\ref{dsigma} that away from the $\bar{D} D^*$ threshold enhancement the $S$-wave component becomes important. In fact, the integrated cross section is dominated by the $S$-wave contributions which will be the dominant source of partial widths of $Y(4260)$ near threshold. But with the increase of the c.m. energy, the $D$-wave contributions will become dominant. This feature can be seen clearly in Fig.~\ref{sigma}. The combined contributions generate the non-trivial cross section lineshape which appear to be consistent with the Belle results although there are still large uncertainties with the data~\cite{Pakhlova:2009jv}.  Future precise measurement of the cross section lineshape at BESIII would be important for confirming this phenomenon.

\begin{table}[htb]
\caption{Parameters determined by fitting the BESIII and Belle experimental data~\cite{{Pakhlova:2009jv},{Ablikim:2013xfr},{Ablikim:2015swa}}.}\label{parameter-list}
\begin{tabular}{|c|c|}
  \hline
  \hline
  Parameters & Fitted values \\
  \hline
  $|y^\text{bare}|$ & $(10.88\pm0.10) \ \text{GeV}^{-1/2}$ \\
  \hline
  $g_1$ & $(29.50\pm0.47) \ \text{GeV}^{-2}$ \\
  \hline
  $1/f_Y$ & $0.063\pm0.011$ \\
  \hline
  $|a|$ & $(12.67\pm0.45) \ \text{GeV}^{-5/2}$ \\
  \hline
  $b$ & $(-15.23\pm0.01) \ \text{GeV}^{2}$ \\
  \hline
  $\chi^2/$d.o.f & 0.92 \\
  \hline \hline
\end{tabular}
\end{table}

The parameters are listed in Table~\ref{parameter-list}. We note that the parameters are strongly constrained by the $\bar{D}D^*+c.c.$ invariant mass spectrum and the recoiled pion angular distributions from BESIII while the data set of cross sections from Belle does not contribute significantly to the $\chi^2$ in the numerical fit. With $Z\equiv 1/[1-\text{Re}(\partial_E\Sigma_1(m_Y))]$ we extract $\alpha^2=Z=0.132\pm0.003$ as the probability of the elementary component inside the $Y(4260)$ and $\beta^2=1-Z=0.868\pm0.003$ is the probability of the $\bar{D}D_1+c.c.$ molecular component. Therefore, the physical wavefunction of $Y(4260)$ is found to be
\begin{equation}\label{wave-2}
|Y(4260)\rangle=0.363|c\bar{c}\rangle+0.932|\bar{D}D_1+c.c.\rangle,
\end{equation}
where we have taken the positive values for both $\alpha$ and $\beta$. Then, the effective couplings can be extracted, i.e. $|y^\text{eff}|=\sqrt{Z}|y^\text{bare}|=(3.94\pm0.04) \ \text{GeV}^{-1/2}$ and $\frac{1}{f^\text{eff}_Y}\leq\sqrt{Z}\frac{1}{f_Y}=0.023\pm0.004$. We then obtain $\Gamma_{Y}^{\text{total}}=(73.0\pm3.5) \ \text{MeV}$ directly from the propagator of $Y(4260)$ at the pole position. Also, the leptonic decay width, $\Gamma(Y(4260)\to e^+e^-)\simeq 483$ eV, can be extracted. This value is consistent with the experimental upper limit~\cite{Ablikim:2015swa}, but significantly larger than the estimate of Ref.~\cite{Dai:2012pb}, i.e. $\Gamma(Y(4260)\to e^+e^-)\simeq 23$ eV. The reason is because that in Ref.~\cite{Dai:2012pb} the contribution from the $\bar{D} D^*\pi +c.c.$ to the total decay width is neglected due to the absence of Breit-Wigner structure in the cross sections of open channels. By assuming that the hidden charm decay channels would saturate the total width of $Y(4260)$, the leptonic decay width will be suppressed. We also mention that our predicted leptonic decay width is also significantly smaller than the recent LQCD estimate~\cite{Chen:2016ejo}, where an upper limit of about 40 eV was set in the hybrid scenario.

In our analysis we find that the open charm channel is still the dominant channel. One interesting feature arising from our scenario is that the partial width for open charm decays is dominated by the $S$-wave decay amplitude which is originated from the HQSS breaking. The $D$-wave decay, which is given by the dominant $\bar{D}D_1$ component in the wavefunction, however, does not contribute largely to the $\bar{D} D^*\pi +c.c.$ channel due to the $D$-wave suppression near threshold. This is similar to the decay of $X(3872)$ which is strongly coupled to $D\bar{D}^*+c.c.$ but has small branching ratio to this channel due to the limited phase space. In this sense, the nontrivial cross section lineshape in the $\bar{D} D^*\pi +c.c.$ channel (see Fig.~\ref{sigma}) would be important to clarify the underlying dynamics. Meanwhile, we emphasize that the wavefunction renormalization parameter $Z$ is found to be $Z=0.132\pm0.003$ which indicates that there exists a large $\bar{D}D_1+c.c.$ component in the $Y(4260)$ wavefunction as shown by Eq.~(\ref{wave-2}). This result is consistent with treating the $Y(4260)$ as
a $\bar{D}D_1+c.c.$ molecule with the mixture of a small compact $c\bar{c}$ component.

With the determination of the leptonic decay width, we can determine the partial widths for those hidden charm decay channels, i.e. $Y(4260)\to J/\psi\pi\pi, \ h_c\pi\pi$ and $\omega\chi_{c0}$. The results are listed in Table~\ref{YtoX-list}. In the $\bar{D} D^*\pi +c.c.$ channel the pion in an $S$ wave or $D$ wave can be separated out and their exclusive contributions are also listed in Table~\ref{YtoX-list}. The sum of these channels gives an estimate of the total width of about 74 MeV. Although this value is smaller than that extracted from $J/\psi\pi\pi$ channel, it should not be surprising since the Breit-Wginer fit of the peak structure in $J/\psi\pi\pi$ can lead to significant uncertainties due to the asymmetric lineshape in the $J/\psi\pi\pi$ channel. A coherent study of different channels within the same framework should be more sensible for understanding the nature of the $Y(4260)$.

\begin{table}[htb]
\caption{The partial widths and branching ratios of $Y(4260)$ determined in the molecular scenario. }\label{YtoX-list}
\center
\begin{tabular}{|c|c|c|}
  \hline
  \hline
  Channels & Width(MeV) & Branching ratio\\
  \hline
  $e^+e^-$ & $(4.83 \pm 1.61) \times 10^{-4} $ & $(4.2 \sim 9.3) \times 10^{-6}$ \\
  \hline
  $(\bar{D}D^*\pi)_{S-\text{wave}}$ & $ 56.54 \pm 4.24 $ & $0.680 \sim 0.875$\\
  \hline
  $(\bar{D}D^*\pi)_{D-\text{wave}}$ & $ 8.37 \pm 0.95 $ & $0.097 \sim 0.134$\\
  \hline
  $\bar{D}D^*\pi$ & $ 64.91\pm 4.45 $ & $0.790 \sim 0.998$\\
  \hline
   $J/\psi \pi\pi$ & $4.68$  & $0.061 \sim 0.067$\\
  \hline
  $h_c \pi \pi$ & $2.67 $   & $ 0.035 \sim 0.038 $  \\
  \hline
  $\omega \chi_{c0}$ & $1.60$  & $ 0.021 \sim 0.023 $  \\
  \hline
   $\Gamma_{Sum}$ & $73.86 \pm 4.45$ & $0.907 \sim 1.1$\\
  \hline \hline
\end{tabular}
\end{table}


\section{The HQSS breaking effects  in the $^3P_0$ model}\label{3P0-model}

By interpreting the $Y(4260)$ to be dominantly a $\bar{D}D_1+c.c.$ molecular state, the relative $S$-wave coupling for $\bar{D}D_1+c.c.$ to an $S$-wave $c\bar{c}$ component is essential for its production in $e^+e^-$ annihilations. As discussed previously that the narrow $\bar{D}D_1+c.c.$ in the HQSS limit will decouple to the $S$-wave $c\bar{c}$~\cite{Li:2013yka}, we will show in this Section that the HQSS breaking will allow a small $S$-wave component to be mixed to the $Y(4260)$ via the $^3P_1$ and $^1P_1$ mixing in the $D_1$ wavefunction.

In the quark model, the physical states $D_1(2420)$ and $D_1(2430)$ are considered as the mixture of the two quark model states $|^1P_1\rangle$ and $|^3P_1\rangle$~\cite{Close:2005se}. Meanwhile, these two states can be written in terms of the spin-parity of the light quark degrees of freedom in language of HQSS, i.e.
\begin{equation}
\left (\begin{array}{c}
|D_1^\prime (2430)\rangle \nonumber\\
|D_1(2420)\rangle\\
\end{array}
\right )=\left (\begin{array}{c}
|1^+,j_l^p=\frac{1}{2}^+\rangle \nonumber\\
|1^+,j_l^p=\frac{3}{2}^+\rangle\\
\end{array}
\right )=\left (\begin{array}{cc}
\cos{\theta}& -\sin{\theta} \nonumber\\
\sin{\theta}& \cos{\theta}\\
\end{array}
\right )\left (\begin{array}{c}
|^1P_1\rangle \nonumber\\
|^3P_1\rangle\\
\end{array}
\right ),
\end{equation}
where in the heavy quark limit the mixing angle $\theta$ takes the ideal mixing angle $\theta_0 = -\arctan(\sqrt{2})=-54.7^\circ$. As the result, the physical state $D_1(2420)$, which is assigned to be the $|1^+,j_l^p=\frac{3}{2}^+\rangle$ state, will decay into $D^*\pi$ via a $D$-wave, thus, becomes narrow. In contrast, the $D_1(2430)$ can decay into $D^*\pi$ via an $S$ wave which makes its width very broad. However, one also recognize that the charm quark mass is not heavy enough to fulfill the heavy quark mass limit of $m_Q\to \infty$. Therefore, some HQSS breaking effects are anticipated.

Adopting the quark model states $|^1P_1\rangle$ and $|^3P_1\rangle$ as the basis states, we can calculate the coupling strength for an $S$-wave $c\bar{c}$ component to the physical $\bar{D}D_1(2420)+c.c.$ The recognition of the HQSS and its breaking can be shown by the coupling dependence on the mixing angle $\theta$. In the
$^3P_0$ model, the general expressions of the partial wave amplitudes for the decay of $1^- \to 1^+(^1P_1)+0^-$ and $1^- \to 1^+(^3P_1)+0^-$ are collected in Table~\ref{pwa-express}.

\begin{table}[htb]
\caption{The general expressions of the partial wave amplitudes for the two-body decay modes of $1^- \to 1^+ + 0^-$ and $1^+ \to 1^- + 0^-$. Here, $\mathcal{I}$ and $\mathcal{F}$ are the relevant isospin and flavor matrix elements, respectively. The spatial function $I_{\pm,0}$ and $T_{\pm,0}$ are presented in the Appendix. \label{pwa-express}}
\begin{tabular}{l l}
  \hline \hline
  \hspace{1cm}Decay mode  & \hspace{3cm} ($J$, $L$) \hspace{4cm} Partial wave amplitudes\hspace{2cm} \\
  \hline
  \hspace{1cm}$1^- \to 1^+(^1 P_1)+0^-$ &\hspace{3cm} (1, 0) \hspace{3cm} $\mathcal{M}_{S-\text{wave}}^{10}=\frac{\sqrt{6}}{9}\gamma \sqrt{E_A E_B E_C} \mathcal{I}\mathcal{F}(I_0+2I_\pm)$\\
  &\hspace{3cm} (1, 2) \hspace{3cm} $\mathcal{M}_{D-\text{wave}}^{12}=\frac{2 \sqrt{3}}{9}\gamma \sqrt{E_A E_B E_C} \mathcal{I}\mathcal{F}(I_\pm-I_0)$\\
  \hline
  \hspace{1cm}$1^- \to 1^+(^3 P_1)+0^-$ &\hspace{3cm} (1, 0) \hspace{3cm} $\mathcal{M}_{S-\text{wave}}^{10}=\frac{2\sqrt{3}}{9}\gamma \sqrt{E_A E_B E_C} \mathcal{I}\mathcal{F}(I_0+2I_\pm)$\\
  &\hspace{3cm} (1, 2) \hspace{3cm} $\mathcal{M}_{D-\text{wave}}^{12}=\frac{\sqrt{6}}{9}\gamma \sqrt{E_A E_B E_C} \mathcal{I}\mathcal{F}(I_0-I_\pm)$\\
 \hline
  \hspace{1cm}$1^+(^1 P_1)\to1^- +0^-$ &\hspace{3cm} (1, 0) \hspace{3cm} $\mathcal{M}_{S-\text{wave}}^{10}=\frac{\sqrt{6}}{9}\gamma \sqrt{E_A E_B E_C} \mathcal{I}\mathcal{F}(T_0-2T_\pm)$\\
  &\hspace{3cm} (1, 2) \hspace{3cm} $\mathcal{M}_{D-\text{wave}}^{12}=-\frac{2 \sqrt{3}}{9}\gamma \sqrt{E_A E_B E_C} \mathcal{I}\mathcal{F}(T_0+T_\pm)$\\
  \hline
  \hspace{1cm}$ 1^+(^3 P_1)\to 1^- +0^-$ &\hspace{3cm} (1, 0) \hspace{3cm} $\mathcal{M}_{S-\text{wave}}^{10}=\frac{2\sqrt{3}}{9}\gamma \sqrt{E_A E_B E_C} \mathcal{I}\mathcal{F}(T_0-2T_\pm)$\\
  &\hspace{3cm} (1, 2) \hspace{3cm} $\mathcal{M}_{D-\text{wave}}^{12}=\frac{\sqrt{6}}{9}\gamma \sqrt{E_A E_B E_C} \mathcal{I}\mathcal{F}(T_0+T_\pm)$\\
  \hline \hline
\end{tabular}
\end{table}

\begin{figure}[htb]
\centering
\scalebox{1.35}{\includegraphics{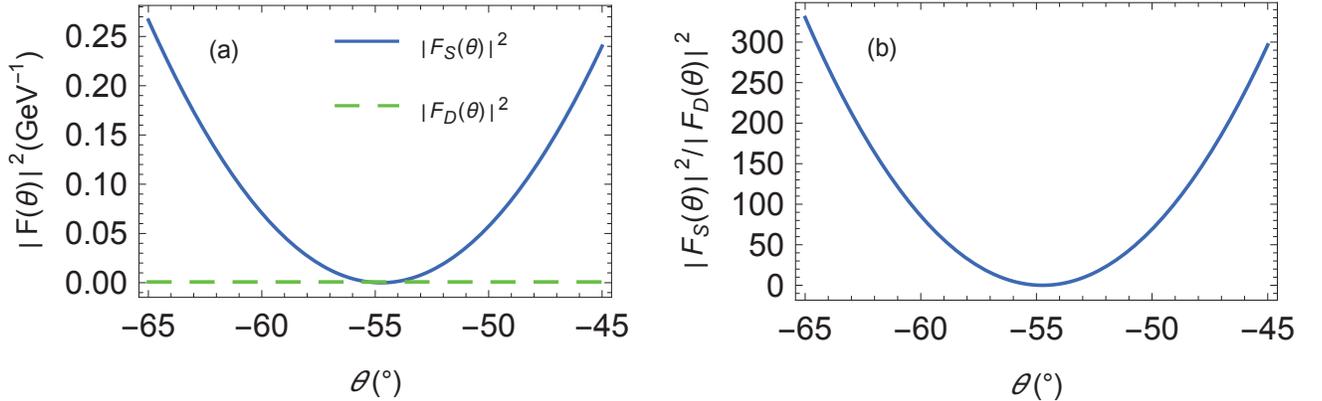}}
\caption{The relative contributions between the $S$ and $D$ wave couplings for $Y(4260)\to \bar{D}D_1$ in terms of the mixing angle $\theta$ in the vicinity of the ideal mixing angle $\theta_0=-54.7^\circ$. Panel (a) denote $|F_S(\theta)|^2$ (solid) and $|F_D(\theta)|^2$ (dashed), while (b) stands for the ratio $|F_S(\theta)|^2/|F_D(\theta)|^2$.} \label{FF-YD1D}
\end{figure}

The bare coupling for $Y(4260)\to \bar{D}D_1+c.c.$ via an $S$ or $D$ wave can then be expressed as
\begin{eqnarray}
\mathcal{M}[Y(4260)\to(\bar{D}D_1+c.c.)_{S-\text{wave}}]&=&\frac{-2(\sin{\theta} + \sqrt{2}\cos{\theta})}{9}\gamma \sqrt{E_A E_B E_C}(I_0+2I_\pm)\equiv \sqrt{8E_A E_B E_C}F_S(\theta), \nonumber \\
\mathcal{M}[Y(4260)\to(\bar{D}D_1+c.c.)_{D-\text{wave}}]&=&\frac{2(\sqrt{2}\sin{\theta} - \cos{\theta})}{9}\gamma \sqrt{E_A E_B E_C}(I_0 - I_\pm)\equiv \sqrt{8E_A E_B E_C}F_D(\theta).
\end{eqnarray}
It can be easily checked that in the HQSS limit with $\theta=\theta_0$ the $S$-wave coupling will vanish as investigated in Refs.~\cite{Li:2013yka,Wang:2013kra,Cleven:2013mka}. Meanwhile, it shows that the $D$-wave coupling for $Y(4260)\to \bar{D}D_1+c.c.$ is highly suppressed. In Fig.~\ref{FF-YD1D} (a), we show the functions $|F_S(\theta)|^2$ (solid line) and $|F_D(\theta)|^2$  (dashed line) in terms of $\theta$ in the vicinity of the ideal mixing angle $\theta_0=-54.7^\circ$. It shows that the $S$-wave coupling increase drastically when $\theta$ deviates from $\theta_0$. This is a direct indication of importance of HQSS breaking effects. In contrast, the $D$-wave coupling negligibly small and its dependence on the mixing angle can also be neglected. In Fig.~\ref{FF-YD1D} (b) the ratio of $|F_S(\theta)|^2/|F_D(\theta)|^2$ is presented. It confirms the drastic increase of the $S$-wave couplings when the HQSS is broken and justifies the strong $S$-wave couplings between $Y(4260)$ and $\bar{D}D_1$. We should mention that since $Y(4260)$ is below the $\bar{D}D_1$ threshold, we adopted a virtual momentum $|\mathbf{K}|=0.234$ GeV in the calculation. This can be regarded as a reasonable estimate of the coupling near threshold.

\begin{figure}[htb]
\centering
\scalebox{1.4}{\includegraphics{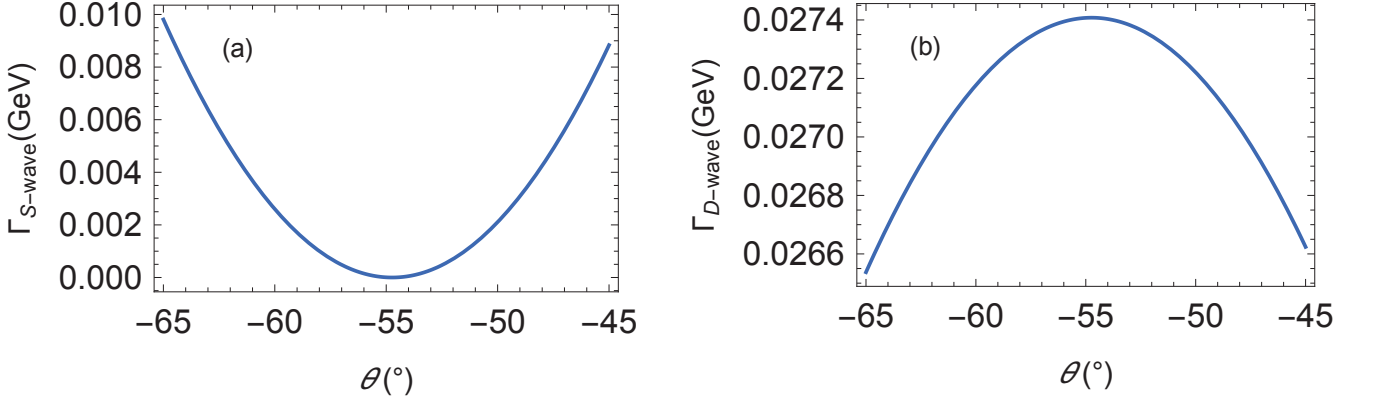}}
\caption{The dependence of the partial decay widths (a) $\Gamma_{S-\text{wave}}$ and (b) $\Gamma_{D-\text{wave}}$ of $D_1^0 \to D^{*+}\pi^-$ on the mixing angle $\theta$.} \label{SD-wave}
\end{figure}

Similarly, we can discuss the strong decay of $D_1(2420)^0 \to D^*(2010)^+\pi^-$ and its HQSS breaking effects. The partial wave amplitudes for its decays into $D^*\pi$ via the $S$ and $D$ wave are also presented in Table.~\ref{pwa-express} and read
\begin{eqnarray}
\mathcal{M}[D_1(2420)^0 \to (D^*(2010)^+\pi^-)_{S-\text{wave}}]&=&\frac{-(\sin{\theta} + \sqrt{2}\cos{\theta})}{9}\gamma \sqrt{E_A E_B E_C}(T_0-2T_\pm), \nonumber \\
\mathcal{M}[D_1(2420)^0 \to (D^*(2010)^+\pi^-)_{D-\text{wave}}]&=&\frac{(\sqrt{2}\sin{\theta} - \cos{\theta})}{9}\gamma \sqrt{E_A E_B E_C}(T_0 + T_\pm).
\end{eqnarray}
where the spatial functions $T_0$ and $T_\pm$ are given in the Appendix. One can see that in the HQSS limit, the $S$-wave decay of $D_1(2420)$ will be suppressed while the $D$-wave decay will be dominant. In Fig.~\ref{SD-wave}, the $S$ and $D$ wave partial widths to $D^*(2010)^+\pi^-$ are presented in terms of the mixing angle in the vicinity of $\theta_0$. It shows that when the mixing angle deviates from the ideal mixing angle it does not change the dominance of the $D$-wave decay. Note that the results of Fig.~\ref{SD-wave} are obtained by assuming that the $D^*\pi$ channel saturates the total width of the $D_1(2420)$, i.e. $\Gamma(D_1\to D^*\pi)=\Gamma_{\text{total}}=(27.4\pm2.5)$ MeV~\cite{Agashe:2014kda}. It gives the $^3P_0$ coupling constant $\gamma=12.17\pm0.54$ which is within a reasonable range of commonly accepted values. But as discussed before, this might have overestimated the coupling for $D_1\to D^*\pi$.

The bare coupling $y^{\text{bare}}$ for $Y(4260)$ to $\bar{D}D_1+c.c.$ in Eq.~(\ref{lagrangian-YD1D}) and coupling $h^\prime$ for $D_1(2420)$ to $D^*\pi$ in Eq.~(\ref{lagrangian-D1}) can then be expressed respectively as
\begin{eqnarray}
|y^{\text{bare}}|&=&\sqrt{8\pi^3}\frac{2\sqrt{2}\gamma |(\sin{\theta} + \sqrt{2}\cos{\theta})I_\pm|}{3}=27.16|(\sin{\theta} + \sqrt{2}\cos{\theta})| \;\text{GeV}^{-1/2} , \nonumber \\
|h^\prime|&=&\sqrt{8\pi^3}\frac{f_\pi}{3\sqrt{6}}\gamma\sqrt{E_\pi}( \chi-\chi^2)\Delta^2|T_\pm|=0.916\pm0.041 \;\text{GeV}^{-1} ,
\end{eqnarray}
where the last equations are determined with $f_\pi=132$ MeV and $\gamma=12.17\pm0.54$. Given that the deviation of the mixing angle from the ideal mixing angle by about $(-5.7\pm 4.0)^\circ$~\cite{Abe:2003zm}, we determine $|y^{\text{bare}}|\simeq 1.36\sim 7.90 \ \text{GeV}^{-1/2}$. This range of value is consistent with the value determined by fitting the experimental data in the previous Section. It allows us to conclude that a small deviation from the HQSS limit can give rise to a strong $S$-wave coupling between a compact $c\bar{c}$ of $1^{--}$ and $\bar{D}D_1(2420)$ and result in a physical state containing large $\bar{D}D_1(2420)+c.c.$ component in the wavefunction.

\begin{table}[htb]
\caption{The helicity amplitudes for the strong decays of $Y(4260) \to \bar{D}D_1(2420)+c.c.$ and $D_1(2420) \to D^*(2010)\pi$. Here, the functions $I_{\pm}$, $T_{\pm}$ and the parameters $\chi$, $\Delta^2$ are listed in the appendix. \label{helicity amplitudes}}
\begin{tabular}{l clcl}
  \hline \hline
  Decay mode &&($M_{J_A}$, $M_{J_B}$, $M_{J_C}$) && Partial wave amplitudes\hspace{2cm} \\
  \hline
  $Y(4260) \to \bar{D}D_1(2420)+c.c.$ && $(1, 1, 0)$ && $\mathcal{M}^{110}=\frac{-(\sin{\theta} + \sqrt{2}\cos{\theta})}{3\sqrt{2}}\gamma \sqrt{8E_A E_B E_C}I_\pm$\\
  && $(0, 0, 0)$ && $\mathcal{M}^{000}=\frac{-(\sin{\theta} + \sqrt{2}\cos{\theta})}{3\sqrt{2}}\gamma \sqrt{8E_A E_B E_C}I_\pm$\\
  &&$(-1, -1, 0)$ && $\mathcal{M}^{-1-10}=\frac{-(\sin{\theta} + \sqrt{2}\cos{\theta})}{3\sqrt{2}}\gamma \sqrt{8E_A E_B E_C}I_\pm$\\
 \hline
 $D_1(2420) \to D^*(2010)\pi$ && $(1, 1, 0)$ && $\mathcal{M}^{110}=\frac{-\cos{\theta}}{12}\gamma \sqrt{8E_A E_B E_C}T_\pm( \chi-\chi^2)\Delta^2\textbf{K}^2$\\
  && $(0, 0, 0)$ && $\mathcal{M}^{000}=\frac{-\sin{\theta}}{6\sqrt{2}}\gamma \sqrt{8E_A E_B E_C}T_\pm( \chi-\chi^2)\Delta^2\textbf{K}^2$\\
  && $(-1, -1, 0)$ && $\mathcal{M}^{-1-10}=\frac{-\cos{\theta}}{12}\gamma \sqrt{8E_A E_B E_C}T_\pm( \chi-\chi^2)\Delta^2\textbf{K}^2$\\
  \hline \hline
\end{tabular}
\end{table}

\section{Summary}\label{summary}

In this work we study the production and decay of $Y(4260)$ in the hadronic molecule picture where the $Y(4260)$ is dominantly a $\bar{D}D_1+c.c.$ molecule with a small compact $c\bar{c}$ component. By combining the constraints from different channels in $e^+e^-$ annihilations, we succeed in describing the available observables simultaneously. Moreover, the invariant mass spectrum for $\bar{D}D^*+c.c.$ in $e^+e^-\to \bar{D} D^*\pi +c.c.$ and the recoiled pion angular distribution to the near-threshold $\bar{D}D^*+c.c.$ peak can be self-consistently accounted for. It shows the importance of the $S$-wave open charm threshold $\bar{D}D_1+c.c.$ in this specific kinematic region for understanding the puzzling phenomena observed in different channels. It also shows that the HQSS breaking plays a crucial role for the production of the $Y(4260)$. We show that a small deviation from the ideal mixing between the HQSS eigenstates will allow sufficient cross sections for $e^+e^-\to \bar{D} D^*\pi +c.c.$ We emphasize that a coherent treatment of the open threshold phenomena could be a key for understanding the puzzling issues with the $Y(4260)$. The future high-statistics analysis at BESIII should be able to clarify the underlying dynamics.

\section*{Acknowledgment}

Useful discussions with M. Cleven, F.-K. Guo, C. Hanhart, Q. Wang, and C.-Z. Yuan are acknowledged. This work is supported, in part, by the National Natural Science Foundation of China (Grant Nos. 11425525 and 11521505),
DFG and NSFC funds to the Sino-German CRC 110 ``Symmetries and the Emergence of Structure in QCD'' (NSFC Grant No. 11261130311), and
National Key Basic Research Program of China under Contract No. 2015CB856700.

\section*{APPENDIX}

The $^3P_0$ quark model~\cite{Micu:1968mk,LeYaouanc:1988fx,Luo:2009wu} has been broadly applied to the study of hadronic transitions.  For simplicity, we only quote the amplitudes for the spatial integrals in $Y(4260)(c\bar{c})\to \bar{D}D_1(2420)+c.c.$ and $D_1\to D^*\pi$ here.

For the decay of $A\to B+C$ for $Y(4260)(c\bar{c})\to \bar{D}D_1(2420)+c.c.$ with the main quantum numbers $n_A=n_B=n_C=0$ and internal orbital angular momenta $L_A=L_C=0$ and $L_B=1$, we have
\begin{eqnarray}
I^{0,m}_{M_{L_B},0}(\mathbf{K})&=&\int\!\rm d^3\mathbf{k}_3 \Psi^*_{0 1 M_{L_B}}\left(\frac{m_3 \mathbf{K}}{m_1+m_3}-\mathbf{k}_3 \right)\Psi^*_{0 0 0}\left(\frac{-m_3 \mathbf{K}}{m_2+m_3}+\mathbf{k}_3 \right)\Psi_{0 0 0}\left(\mathbf{K}-\mathbf{k}_3 \right)\mathcal{Y}_{1m}(\mathbf{k}_3)\nonumber\\
&=&\frac{i \sqrt{2}}{\pi^{9/4}}\sqrt{\frac{3}{4\pi}}R_A^{3/2}R_B^{5/2}R_C^{3/2} \exp(-\frac{1}{2} \xi^2 \mathbf{K}^2)\int \rm d^3 \mathbf{k} \left(\mathbf{k}_{M_{L_B}}^* \mathbf{k}_m+(\beta\eta+\eta^2)\mathbf{K}_{M_{L_B}}^* \mathbf{K}_m\right)\exp(-\frac{1}{2} \Delta^2 \mathbf{k}^2),\nonumber
\end{eqnarray}
where $\Delta^2\equiv R_A^2+R_B^2+R_C^2$, $\xi^2\equiv \alpha^2R_A^2+\delta^2R^2_C-\frac{(\alpha R_A^2+\delta R^2_C)^2}{R_A^2+R_B^2+R_C^2}$, $\alpha\equiv \frac{m_1}{m_1+m_3}$, $\beta\equiv \frac{m_3}{m_1+m_3}$ , $\delta\equiv (\frac{m_3}{m_2+m_3}-\frac{m_3}{m_1+m_3})$, $\eta\equiv \frac{\alpha R_A^2+\delta R^2_C}{R_A^2+R_B^2+R_C^2}$, with $k_{\pm1}\equiv\mp(k_x\pm k_y)/\sqrt{2}$ and $k_0\equiv k_z$.

By choosing the direction of $\mathbf{K}$ as the $z$ axis, we obtain
\begin{eqnarray}\label{eq-Y-decay}
&&I_{\pm}\equiv I_{1,0}^{0,1}=I_{-1,0}^{0,-1}=i\frac{2\sqrt{3}}{\pi^{5/4}\Delta^5}
\left(R_A^{3/2}R_B^{5/2}R_C^{3/2}\right)\exp\left(-\frac{1}{2}\xi^2\textbf{K}^2\right),\nonumber\\
&&I_0\equiv I_{0,0}^{0,0}=i\frac{2\sqrt{3}}{\pi^{5/4}\Delta^5}\left(R_A^{3/2}R_B^{5/2}R_C^{3/2}\right)\exp\left(-\frac{1}{2}\xi^2\textbf{K}^2\right)\left[1+(\beta\eta +\eta^2)\Delta^2\mathbf{K}^2\right],
\end{eqnarray}
while in other cases, we have $I_{M_{L_B},0}^{0,m}=0$.

Similarly, for $D_1\to D^*\pi$ with $n_A=n_B=n_C=0$, $L_A=1$ and $L_B=L_C=0$, we have
\begin{eqnarray}
I^{M_{L_A},m}_{0\,,\; 0}(\mathbf{K})&=&\int\!\rm d^3\mathbf{k}_3 \Psi^*_{000}\left(\frac{m_3 \mathbf{K}}{m_1+m_3}-\mathbf{k}_3 \right)\Psi^*_{0 0 0}\left(\frac{-m_3 \mathbf{K}}{m_2+m_3}+\mathbf{k}_3 \right)\Psi_{0 1 M_{L_A}}\left(\mathbf{K}-\mathbf{k}_3 \right)\mathcal{Y}_{1m}(\mathbf{k}_3)\nonumber \\
&=&\frac{i \sqrt{2}}{\pi^{9/4}}\sqrt{\frac{3}{4\pi}}R_A^{5/2}R_B^{3/2}R_C^{3/2} \exp(-\frac{1}{2} \zeta^2 \mathbf{K}^2)\int \rm d^3 \mathbf{k} \left(-\mathbf{k}_{M_{L_A}} \mathbf{k}_m+(\chi-\chi^2)\mathbf{K}_{M_{L_A}} \mathbf{K}_m\right)\exp(-\frac{1}{2} \Delta^2 \mathbf{k}^2),\nonumber
\end{eqnarray}
where $\Delta^2\equiv R_A^2+R_B^2+R_C^2$, $\zeta^2\equiv R_A^2+\beta^2R_B^2+\gamma^2R_C^2-\frac{(R_A^2+\beta R_B^2+\gamma R_C^2)^2}{R_A^2+R_B^2+R_C^2}$, $\chi\equiv \frac{R_A^2+\beta R_B^2+\gamma R_C^2}{R_A^2+R_B^2+R_C^2}$, $\beta\equiv \frac{m_3}{m_1+m_3}$, and $\gamma\equiv \frac{m_3}{m_2+m_3}$.

By choosing the direction of $\mathbf{K}$ as the $z$ axis, the explicit spatial integral gives
\begin{eqnarray}\label{D1-decay}
&&T_{\pm}\equiv I^{1,-1}_{0,0}=I^{-1,1}_{0,0}=i\frac{2\sqrt{3}}{\pi^{5/4}\Delta^5}
\left(R_A^{5/2}R_B^{3/2}R_C^{3/2}\right)\exp\left(-\frac{1}{2}\zeta^2\textbf{K}^2\right),\nonumber\\
&&T_0\equiv I_{0,0}^{0,0}=i\frac{2\sqrt{3}}{\pi^{5/4}\Delta^5}\left(R_A^{5/2}R_B^{3/2}R_C^{3/2}\right)\exp\left(-\frac{1}{2}\zeta^2\textbf{K}^2\right)\left[-1+( \chi-\chi^2)\Delta^2\textbf{K}^2\right],
\end{eqnarray}
while in other cases, we have  $I^{M_{L_A},m}_{0\,,\; 0}(\mathbf{K})=0$.

We note the in Eqs.~(\ref{eq-Y-decay}) and (\ref{D1-decay}) the harmonic oscillator strengths $R_A$, $R_B$ and $R_C$ have values according to the interacting states. Also, the kinematic variables are defined by the processes. The harmonic oscillator strengths for different states are listed in Table~\ref{para}. The following constitute quark masses are adopted in the calculation, i.e. $m_c=1.55$ GeV, and $m_u=m_d=0.34$ GeV.

\begin{table}[htb]
\caption{Harmonic oscillator strengths~\cite{Godfrey:1986wj} for different particles involved in $Y(4260)(c\bar{c}) \to \bar{D} D_1+c.c.$ and $D_1(4260) \to D^*\pi$. \label{para}}
\begin{tabular}{l c c c c c }
  \hline
  \hline
  & $Y(4260)(c\bar{c})$  & $D_1(2420)$  & $ D$  & $D^*(2010)$  & $\pi$  \\
%  \hline
%  Mass (GeV)& 4.217 & 2.422 & 1.864 & 2.010 & 0.140 \\
  \hline
  $R$ ($\text{GeV}^{-1}$) & 1.52 & 3.03 & 2.33 & 2.70 & 2.5 \\
  \hline \hline
\end{tabular}
\end{table}

\end{document}